\documentclass[11pt]{amsart}
\pdfoutput=1
\usepackage{times}
\usepackage{comment}
\usepackage{epsfig}
\usepackage{varioref}
\usepackage{textcomp}
\usepackage{multirow}
\usepackage{graphics}
\usepackage{bmpsize}
\usepackage{url}
\usepackage{draftcopy}
\usepackage[numbers]{natbib}
\usepackage{amssymb}
\usepackage{algorithm}
\usepackage{algpseudocode}
\usepackage{varioref}
\usepackage{multirow}
\usepackage{enumitem}
\makeatletter\@ifclassloaded{beamer}{}{\usepackage{imakeidx}}\makeatother
\usepackage{lscape}
\usepackage{gensymb}
\usepackage[toc]{appendix}
\usepackage{mathtools}
\usepackage{pict2e}
\usepackage{picture}
\newcommand{\indexed}[1]{#1\index{#1}}

\DeclareMathOperator*{\argmax}{arg~max}
\DeclareMathOperator*{\argmin}{arg~min}

\DeclareMathOperator*{\where}{where}

\DeclareMathOperator*{\sign}{sgn}

\makeatletter\@ifclassloaded{beamer}{}{}\makeatother

\newcommand{\infinity}{\infty}

\newcommand{\genlab}[2]{\label{#1:#2}}
\newcommand{\genref}[2]{#1~\vref{#1:#2}}

\newcommand{\seclab}[1]{\genlab{section}{#1}}

\newcommand{\eqnlab}[1]{\genlab{equation}{#1}}
\newcommand{\eqnref}[1]{\genref{equation}{#1}}

\newcommand{\figlab}[1]{\genlab{figure}{#1}}
\newcommand{\figref}[1]{\genref{figure}{#1}}

\makeatletter\@ifclassloaded{beamer}{}{%

}
\makeatother
\newcommand{\pdffigure}[4]{\begin{center}\begin{figure}[hbt]\includegraphics[angle=#4,scale=#3]{#1.pdf}\caption[\hspace{\normalparindent}#2]{#2}\figlab{#1}\end{figure}\end{center}}
\makeatletter\@ifclassloaded{beamer}{\input{beamer}}{}\makeatother

\vrefwarning
\begin{document}
%
%
\title{An Analytic Solution for Asset Allocation with a Multivariate Laplace Distribution}
\author{Graham L. Giller}
\email{graham@gillerinvestments.com}
\date{\today}
\begin{abstract}
In this short note the theory for multi-variate asset allocation with elliptically symmetric distributions of returns, as developed in the authors prior work, is specialized to the case of returns drawn from a multi-variate Laplace distribution. This analysis delivers a result closely, but not perfectly, consistent with the conjecture presented in the author's article \textit{Thinking Differently About Asset Allocation}. The principal differences are due to the introduction of a term in the dimensionality of the problem, which was omitted from the conjectured solution, and a re-scaling of the variance due to varying parameterizations of the univariate Laplace distribution.
\end{abstract}
\maketitle
\section{Asset Allocation with Elliptically Symmetric Distributions}
\subsection{A Result for Negative Exponential Utility Maximizers}
In the authors prior works\cite{giller2004frictionless,giller2022adventures,giller2023essays}, a solution for multivariate asset allocation with elliptically symmetrical distributions of returns is presented. In terms of a ``holding function,'' which is a policy function that maps the conditional moments of a distribution of future returns known to the trader to the optimal position they should hold in the assets described by that returns distribution, it is shown to be
\begin{equation}\eqnlab{ellipsoidal}
\hat{\boldsymbol{h}}_t=\hat{\boldsymbol{h}}(\boldsymbol{\alpha}_t,\Sigma_t)
=\frac{\Sigma_t^{-1}\boldsymbol{\alpha}_t}{\lambda\Psi_\frac{n}{2}\!(\hat{x}_t)}.
\end{equation}
This applies for a distribution of returns which may be written $f(g^2)$ where
\begin{equation}\eqnlab{mahanalobis}
    g^2=(\boldsymbol{r}_t-\boldsymbol{\alpha}_t)^T\Sigma_t^{-1}(\boldsymbol{r}_t-\boldsymbol{\alpha}_t)
\end{equation}
is the Mahanalobis distance\cite{mardia1979} between the return, $\boldsymbol{r}_t$, and it's central tendency, $\boldsymbol{\alpha}_t$, scaled by a matrix, $\Sigma_t$, that is proportional to the covariance matrix of returns, $V_t$, in a distribution dependent manner. This expression is derived for a maximizer of negative exponential utility and, in the references, the ``scaling function'' $\Psi_\nu(x)$ is defined by
\begin{equation}\eqnlab{psifun}
\Psi_\nu(x)=\frac{1}{x}\frac{\int_0^\infinity f(g^2)I_\nu(gx)\,g^{\nu+1}\,dg}{\int_0^\infinity f(g^2)I_{\nu-1}(gx)\,g^\nu\,dg}.
\end{equation}
Here $I_\nu(\cdot)$ is the modified Bessel function of the first kind\cite{abramowitz258}. For $n$ assets, the \textit{critical root}, $\hat{x}_t$, is defined to be the solution of
\begin{equation}\eqnlab{xsolve}
\hat{x}_t=x\;:\;x\Psi_\frac{n}{2}(x)=\sqrt{\boldsymbol{\alpha}_t^T\Sigma^{-1}_t\boldsymbol{\alpha}_t}.
\end{equation}

$\Psi_\nu(x)$ is a monotonic increasing function of its non-negative argument, $x$, that acts to ``scale down'' positions in the presence of ``fat tails,'' i.e. in situations where the distribution of returns, $f$, has significant excess kurtosis. For the Normal distribution it is equal to unity for all values of it's argument.
\subsection{The Generalized Error Distribution}
In the majority of the author's prior work, the distribution of returns has been taken to be the \textit{Generalized Error distribution}, parameterized as 
\begin{equation}\eqnlab{multipdf}
f(\boldsymbol{r}_t|\boldsymbol{\alpha}_t,\Sigma_t,\kappa)=
\frac{1}{\sqrt{\pi^n|\Sigma_t|}}
\frac{\Gamma(1+\frac{n}{2})}{\Gamma(1+n\kappa)}
\left\{\frac{\Gamma(3\kappa)}{\Gamma(\kappa)}\right\}^\frac{n}{2}
e^{-\left\{\frac{\Gamma(3\kappa)}{\Gamma(\kappa)}
g^2
\right\}^\frac{1}{2\kappa}}.
\end{equation}
The covariance matrix for this distribution exists, and is a scaling of the parameter matrix $\Sigma_t$:
\begin{equation}\eqnlab{vsigma}
V_t=\mathbb{V}[\boldsymbol{r}_t]=\frac{\Gamma\{(n+2)\kappa\}\Gamma(1+\kappa)}{\Gamma(3\kappa)\Gamma(1+n\kappa)}\Sigma_t.
\end{equation}
Note that both the dimensionality, $n$, and the kurtosis parameter, $\kappa$, affect the scaling between the covariance matrix, $V_t$, and the matrix, $\Sigma_t$.

This particular parameterization is chosen \textit{entirely} because it is identical to the multinormal distribution when $\kappa=1/2$. In that limit the covariance matrix is simply $\Sigma_t$. Thus it is possible to use this form to perform a composite hypothesis test for multi-normality of asset returns. Numerical solutions to \eqnref{ellipsoidal} for the case of a Generalized Error distribution with $\kappa\approx0.75$, which is the value supported by the author's empirical investigations\cite{giller2022normal}, are presented in references\cite{giller2004frictionless} and \cite{giller2022adventures,giller2023essays}.
\subsection{A Multivariate Laplace Distribution}
The form of \eqnref{multipdf} also encompasses a multivariate Laplace distribution when $\kappa=1$: in that case
\begin{equation}\eqnlab{multilaplace}
f(\boldsymbol{r}_t|\boldsymbol{\alpha}_t,\Sigma_t)=
\sqrt{\frac{2^n}{\pi^n|\Sigma_t|}}
\frac{\Gamma(1+\frac{n}{2})}{\Gamma(1+n)}
e^{-\sqrt{2(\boldsymbol{r}_t-\boldsymbol{\alpha}_t)^T\Sigma_t^{-1}(\boldsymbol{r}_t-\boldsymbol{\alpha}_t)}}.
\end{equation}
This is \textit{not} the only manner in which a multivariate generalization to the Laplace distribution might be defined; however, it is one which possesses elliptical symmetry. An important feature of \textit{this} form is that the marginal univariate distributions are \textit{not} themselves univariate Laplace distributions\cite{forbes2011statistical}. With this form, it is only the \textit{radial} coordinate, written $g$ in \eqnref{mahanalobis}, that has a marginal Laplace distribution.
\section{An Analytic Solution for the Univariate Laplace Distribution}
Conventionally, the univariate Laplace distribution is written with the following parameterization\footnote{As adapted to support the current context of an active trader.}
\begin{equation}\eqnlab{laplacepdf}
f(r_t|\alpha_t,\sigma_t)=\frac{1}{2\sigma_t}e^{-\left|\frac{r_t-\alpha_t}{\sigma_t}\right|}.
\end{equation}
This distribution has a mean $\alpha_t$ and variance $2\sigma_t^2$. A negative exponential utility maximizer seeks to find the solution to
\begin{equation}\eqnlab{laplaceutility}
\hat{h}_t=\argmin_{h_t}\int_{-\infty}^\infty\frac{1}{2\sigma_t}e^{-\lambda h_tr_t-\left|\frac{r_t-\alpha_t}{\sigma_t}\right|}\,dr_t,
\end{equation}
where $\lambda$ represents the ``price of risk,'' or the rate at which a given trader will exchange expected profits for expected variance in their profits. As discussed in Giller\cite{giller2023essays}, the integral may be evaluated analytically as
\begin{equation}\eqnlab{laplaceomega}
\omega(h_t)=\frac{e^{-\lambda h_t\alpha_t}}{1-\lambda^2h_t^2\sigma_t^2}\;\mbox{where}\;|\lambda h_t\sigma_t|<1.
\end{equation}
Differentiating w.r.t.\ $h_t$ and solving for $d\omega/dh_t=0$ gives two roots, both of which are increasing functions of the expected return, $\alpha_t$. However, only one root, which can be labeled the ``positive solution,'' also satisfies the condition that $\sign h_t=\sign\alpha_t$. 

This is the solution that maximizes utility\index{utility!maximization}, and so the ``\indexed{holding function}''\footnote{i.e.\ The optimal policy function for a negative exponential utility maximizing trader.} for the \indexed{Laplace distribution} is
\begin{equation}\eqnlab{laplaceholding}
h(\alpha_t,\sigma_t)=\frac{\sqrt{1+\alpha_t^2/\sigma_t^2}-1}{\lambda\alpha_t}.
\end{equation}
In terms of the alpha, this holding function has the Taylor series
\begin{equation}\eqnlab{laplacetaylor}
h(\alpha_t,\sigma_t)=\frac{\alpha_t}{2 \lambda  \sigma_t^2}-\frac{\alpha_t^3}{8\lambda\sigma_t^4}+O\left(\alpha_t^5\right),
\end{equation}
which recovers the solution to the univariate version of Markowitz's mean-variance optimization problem\cite{markowitz1991founders} for small alpha, but has the limit
\begin{equation}\eqnlab{laplacelimit}
\lim_{\alpha_t\rightarrow\pm\infty}h(\alpha_t,\sigma_t)=\pm\frac{1}{\lambda\sigma_t}
\end{equation}
for large alphas. 

The author understands that this solution is not well known of in the quantitative finance community\cite{kyle2023private}. In prior work, numerical solutions to equations \ref{equation:psifun} and \ref{equation:xsolve} are found to exhibit very similar scaling behaviour to that observed in \eqnref{laplaceholding} for large alphas in the case of a Generalized Error distribution with $\kappa\approx0.75$, which seems to be supported by empirical work\cite{giller2022normal}. Prior to this work, a similar investigation had not been executed for a multivariate Laplace distribution such as that of \eqnref{multilaplace}.
\section{The Prior Conjecture}\seclab{conjecture}
Aware of the results of both \eqnref{ellipsoidal} and \eqnref{laplaceholding}, the author conjectured in Giller\cite{giller2023thinking}, that \textit{all} solutions to multivariate optimal trading strategies take the form\footnote{It might be more reasonable to restrict this conjecture to distributions with ellipsoidal symmetry.}
\begin{equation}\eqnlab{conjecture}
    \boldsymbol{h}(\boldsymbol{\alpha}_t,V_t)=\frac{V_t^{-1}\boldsymbol{\alpha_t}}{2\lambda}\Omega(Z_t),
\end{equation}
where $Z_t$ is the Mahanalobis distance of the alpha from its central tendency, $\boldsymbol{0}$, scaled by the covariance matrix of returns, $V_t$. i.e.
\begin{equation}\eqnlab{mananalobisv}
    Z_t=\sqrt{\boldsymbol{\alpha_t}^TV_t^{-1}\boldsymbol{\alpha_t}}.
\end{equation}
which can be though of as a ``$Z$-score'' for the alpha. The function $\Omega(Z_t)$ takes the place of $1/\Psi_{n/2}(\hat{x}_t)$ in equation \ref{equation:ellipsoidal} for this conjectured generalization of the result. This $\Omega(Z_t)$ is a strictly positive \textit{decreasing} function of it's argument that takes the value $1$ at the origin and is identically equal to that value for all values of the argument in the case of the Normal distribution. It plays the similar role of decreasing traders bets on ``big'' alphas in markets that have fat tails (leptokurtosis).
\subsection{The Univariate Laplace Solution in the Conjectured Form}
The result of \eqnref{laplaceholding} is not immediately in the conjectured form, but this can be delivered via simple algebraic identities. For the univariate solution let $Z_t=|\alpha_t/\sigma_t|$, then
\begin{align}
    \frac{1}{\alpha_t}&=\frac{\alpha_t}{\alpha_t^2}=\frac{\alpha_t/\sigma_t^2}{\alpha_t^2/\sigma_t^2}
    =\frac{\alpha_t/\sigma_t^2}{Z_t^2}\\
    \Rightarrow h(\alpha_t,\sigma_t)&=\frac{\alpha_t/\sigma_t^2}{\lambda}
    \frac{\sqrt{1+Z_t^2}-1}{Z_t^2}\eqnlab{laplaceconj}\\
    &=\frac{\alpha_t/\sigma_t^2}{2\lambda}\Omega(Z_t)\\
    \mathrm{where}\;\Omega(Z_t)&=2\frac{\sqrt{1+Z_t^2}-1}{Z_t^2}
\end{align}
\subsection{The Conjectured Multivariate Laplace Solution}
The transformation of \eqnref{laplaceholding} to the conjectured form as exhibited in \eqnref{laplaceconj} then follows immediately from dimensional arguments:
\begin{align}
    \alpha_t/\sigma_t^2&\rightarrow V^{-1}_t\boldsymbol{\alpha_t}\\
    Z_t^2&\rightarrow\boldsymbol{\alpha}_t^TV_t^{-1}\boldsymbol{\alpha}_t\\
    \Rightarrow h(\boldsymbol{\alpha}_t,V_t)&=\frac{V_t^{-1}\boldsymbol{\alpha_t}}{2\lambda}\times2\frac{\sqrt{1+Z_t^2}-1}{Z_t^2}\eqnlab{mvlconj}
\end{align}
\section{An Analytic Solution for the Multivariate Laplace Distribution}
Notwithstanding the results presented above, an analytical solution for the multi-variate Laplace distribution of \eqnref{multilaplace} will now be developed.
\subsection{The Scaling Function}
For the form of the Generalized Error distribution given in \eqnref{multipdf} the scaling function becomes
\begin{equation}\eqnlab{gedpsi}
\Psi_\nu(x)=\frac{1}{x}\frac{\int_0^\infinity e^{-\eta g^\frac{1}{k}}I_\nu(gx)g^{\nu+1}\,dg}
{\int_0^\infinity e^{-\eta g^\frac{1}{k}}I_{\nu-1}(gx)g^\nu\,dg}
\,\where\,\eta=\left\{\frac{\Gamma(3\kappa)}{\Gamma(\kappa)}\right\}^\frac{1}{2\kappa}.
\end{equation}
For the specific case of the multivariate Laplace distribution given here, with $\kappa=1$, this becomes
\begin{equation}\eqnlab{gedpsi1}
\Psi_\nu(x)=\frac{1}{x}\frac{\int_0^\infinity e^{-\sqrt{2}g}I_\nu(gx)g^{\nu+1}\,dg}
{\int_0^\infinity e^{-\sqrt{2}g}I_{\nu-1}(gx)g^\nu\,dg}
\,\mathrm{since}\,\eta=\sqrt{2}.
\end{equation}
In \eqnref{gedpsi1}, the integral in the numerator evaluates to
\begin{equation}\eqnlab{n}
\frac{2^{\nu +\frac{3}{2}} \Gamma \left(\nu +\frac{3}{2}\right) x^{\nu }
   \left(\frac{1}{2-x^2}\right)^{\nu +\frac{3}{2}}}{\sqrt{\pi }}
\end{equation}
and that in the denominator to
\begin{equation}\eqnlab{d}
\frac{2^{\nu +\frac{1}{2}} \Gamma \left(\nu +\frac{1}{2}\right) x^{\nu -1}
   \left(\frac{1}{2-x^2}\right)^{\nu +\frac{1}{2}}}{\sqrt{\pi }}.
\end{equation}
The condition $x<\eta=\sqrt{2}$ is neccesary to guarantee convergence of the integrals. In \eqnref{gedpsi1} the majority of terms from the integral cancel, leaving a remarkably simple result
\begin{equation}\eqnlab{getspi2}
\Psi_\nu(x)=\frac{1+2\nu}{2-x^2}\;\mathrm{or}\;\Psi_{\!\frac{n}{2}}(x)=\frac{1+n}{2-x^2}.
\end{equation}
\subsection{The Critical Root and the Optimal Holding Function}
The critical root, $\hat{x}_t$, is then
\begin{equation}
    \hat{x}_t=\frac{\sqrt{(n+1)^2+8 {Z'_t}^2}-(n+1)}{2 Z'_t}, 
\end{equation}
where ${Z'_t}^2=\boldsymbol{\alpha}_t\Sigma_t^{-1}\boldsymbol{\alpha_t}$ is the Manahalobis distance in terms of the scaling matrix, $\Sigma$. This gives
\begin{equation}
    \boldsymbol{h}\eqnlab{mvlexact}(\boldsymbol{\alpha}_t,\Sigma_t)=\frac{\Sigma_t^{-1}\boldsymbol{\alpha_t}}{\lambda}\frac{\sqrt{\left(\frac{n+1}{2}\right)^2+2 {Z'_t}^2}-\frac{n+1}{2}}{{Z'_t}^2}
\end{equation}
for the optimal portfolio.

Now, for this parameterization of a multivariate Laplace distribution, the relationship between the covariance matrix and the scaling matrix, using \eqnref{vsigma}, is just
\begin{equation}
    V_t=\frac{n+1}{2}\Sigma_t
\end{equation}
which means that we can re-express $Z'_t$ in terms of the ``more traditional'' Mahanalobis distance using the covariance matrix, $V_t$:
\begin{equation}\eqnlab{zprimez}
    Z_t^2=\boldsymbol{\alpha}_tV^{-1}_t\boldsymbol{\alpha}_t
         =\frac{2}{n+1}\boldsymbol{\alpha}_t\Sigma^{-1}_t\boldsymbol{\alpha}_t
         =\frac{2}{n+1}{Z'_t}^2.
\end{equation}
Substituting this relationship into \eqnref{mvlexact} gives
\begin{equation}\eqnlab{mvlexact2}
\boldsymbol{h}(\boldsymbol{\alpha}_t,V_t)=\frac{ V_t^{-1}\boldsymbol{\alpha_t}}{2\lambda}\frac{\sqrt{1+4Z_t^2/(n+1)}-1}{Z_t^2/(n+1)}   
\end{equation}
or
\begin{equation}\eqnlab{omegaexact}
\Omega_n(Z_t)=2\frac{\sqrt{1+4 Z_t^2/(n+1)}-1}{2Z_t^2/(n+1)}. 
\end{equation}
\subsection{Comparison of the Analytic Solution with the Conjectured Form}
The form of \eqnref{mvlexact} differs slightly from the conjectured form of \eqnref{mvlconj}. Most importantly, the dimensionality of the problem is exhibited through the term $(n+1)/2$ which becomes unity for the univariate case and is seen in \eqnref{laplaceholding} but not within the conjectured form. This is an error in the conjecture.

Secondly, there is a factor of two within the root preceding $Z_t^2$. This, in fact, arises from the differing parameterization of the Laplace distribution given. The conventional form, given in \eqnref{laplacepdf} has a variance of $2\sigma_t^2$, whereas the univariate limit of the form given in \eqnref{multilaplace} is, in fact,
\begin{equation}
f(r_t|\alpha_t,\sigma_t)=\frac{1}{\sqrt{2}\sigma_t}e^{-\sqrt{2}\left|\frac{r_t-\alpha_t}{\sigma_t}\right|}
\end{equation}
which has a variance of exactly $\sigma_t^2$. Thus this ``error'' in the conjecture is largely trivial.
\subsection{The Large Portfolio Limit}
The shape of the $\Omega_n(Z_t)$ function for various values of $n$ is illustrated in \figref{omega}. From the figure can be seen a symmetric profile that strongly disfavours alphas with high values of $Z_t$ in a univariate portfolio, but that this down-weighting is diffused for larger portfolios.
\pdffigure{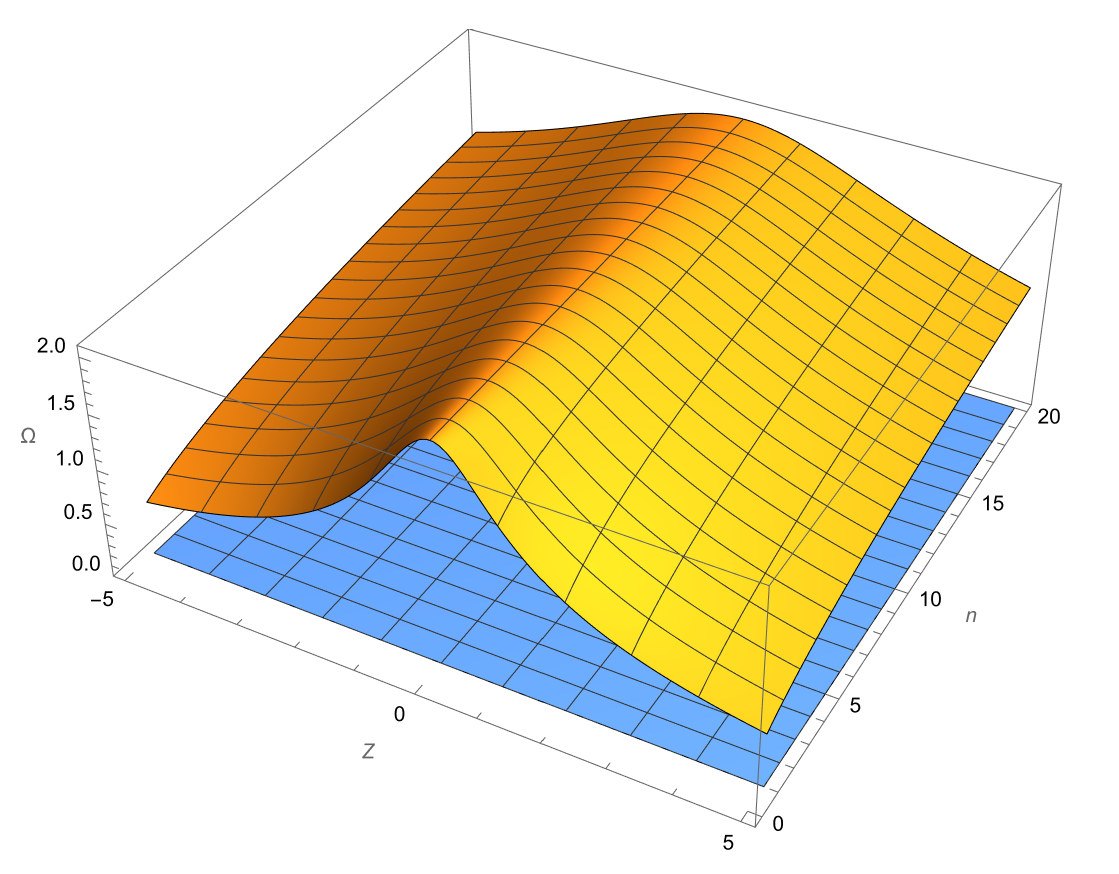}{The $\Omega_n(Z_t)$ function for a multivariate Laplace distribution.}{0.5,bb=0 0 6.95in 5.54in}{0}

Although it is tempting to examine the unconditional limit of very large portfolios 
\begin{equation}
\lim_{n\rightarrow\infty}\Omega_n(Z_t)=2   
\end{equation}
this is not, in fact, the correct limit as $Z_t$ is, in general, a stochastic quantity and $\mathbb{E}[Z_t^2]$ is $o(n)$ for a portfolio of size $n$ (and exactly that for Normally distributed asset returns). In contrast
\begin{equation}
\lim_{n\rightarrow\infty}\Omega_n(\sqrt{n})=\sqrt{5}-1.
\end{equation}
and, since the maximum value\footnote{Note that a factor of two is somewhat arbitrarily inserted into \eqnref{mvlconj} for consistency with the canonical Markowitz solution. Without that, the maximum value of $\Omega(Z_t)$ would be one. The work presented here suggests that it might be a bit more ``natural'' to replace $\lambda$ with $\lambda/2$ in the Mean Variance Optimization framework, which would eliminate this factor.} of $\Omega(Z_t)$ is $2$ for all $n$, it is amusing\footnote{This value is not regarded as \textit{significant} by the author.} to note that
\begin{equation}\eqnlab{golden}
\lim_{n\rightarrow\infty}\frac{\Omega_n(\sqrt{n})}{2}=\frac{1}{\varphi}
\end{equation}
where $\varphi$ is the golden ratio.

The limit of \eqnref{golden} is restricted to the point $Z_t^2=n$. Replacing this with a more general expression of $Z_t^2=\zeta^2n$, where $\zeta=o(1)$, gives
\begin{equation}
\lim_{n\rightarrow\infty}\Omega_n(\zeta\sqrt{n})=\frac{\sqrt{1+4\zeta^2}-1}{\zeta^2}.
\end{equation}
\subsection{Asymptotic Behaviour of the Scaling Function}
Stepping back from the infinite limit of the portfolio scaling function, it is informative to examine it's asymptotic properties as the arguments become large. This is easy to compute: \eqnref{omegaexact} it's clear that
\begin{equation}
    \lim_{Z_t^2/(n+1)\rightarrow\infty}\Omega_n(Z_t)\sim\frac{2}{|Z_t|/\sqrt{n+1}}.
\end{equation}
\subsection{Elimination of the Lagrange Multiplier}
An aspect of the mean-variance optimization procedure that is problematic for analysts seeking a ``complete'' solution to the asset allocation problem is the continued presence of the Lagrange multiplier, $\lambda$, in the solution. One way to eliminate this term is to use constraints to fix it to a value that derives from those constraints.

For example, it is common to write the m.v.o.\ problem as\footnote{With $\boldsymbol{1}$ as the ``unit'' vector $\boldsymbol{1}^T=(1\;1\dots 1)$.}
\begin{equation}
\hat{\boldsymbol{h}}_t=\argmax_{\boldsymbol{h}_t}\left(\boldsymbol{h}_t^T\boldsymbol{\alpha}_t-\lambda\boldsymbol{h}_t^TV_t^{-1}\boldsymbol{h}_t\right.
    \left|\;\boldsymbol{h}_t^T\boldsymbol{1}=1\right).
\end{equation}
This refined problem represents mean-variance optimization with a ``total net investment'' constraint and the solution, $\hat{\boldsymbol{h}}_t$, represents the proportion of the portfolio to invest into each asset. Solution is straightforward, and the specific value of the Lagrange multiplier is set by the constraint, giving:
\begin{equation}\eqnlab{markowitz1}
    \hat{\boldsymbol{h}}_t=\frac{V_t^{-1}\boldsymbol{\alpha}_t}{\boldsymbol{1}^TV_t^{-1}\boldsymbol{\alpha}_t}.
\end{equation}
An important feature of this portfolio is that it is not proportional to expected returns and not inversely proportional to risk. It is not a ``gains seeking, risk averse'' allocation at all as it will always be maximally exposed to risk. The only thing that is affected by the specific values of the alpha and the covariance matrix is the ``direction'' in the solution space, $\mathbb{R}^n$, favoured, since \eqnref{markowitz1} is a unit-vector in that space by virtue of the applied constraint.

Applying the same construction to \eqnref{mvlexact2} delivers exactly the same result, that the optimal portfolio is given by \eqnref{markowitz1} and that the chosen portfolio does not have a scale that is dependent on any risk-reward trade-off, or moderated by a factor, $\Omega_n(Z_t)$, that arises from the kurtosis of the returns. Thus it can be seen that an investors insistence of full investment, no matter how small the return is relative to the risk, is the factor that leads to this result and that choice is also a choice to ignore the effects of the kurtosis of the distribution of returns on the optimal portfolio chosen. If the investor wishes to be averse to the higher moments of the distribution of portfolio returns, they should avoid portfolio constraints that immunize them to their aversion to such moments.
\section{Conclusions}
In this brief note three important results are presented:
\begin{enumerate}
    \item a full analytic solution for asset allocation strategy that should be implemented by a negative exponential utility maximizer for a single asset when the returns of that asset are drawn from a univariate Laplace distribution;
    \item a full analytic solution for an equivalent problem when returns are drawn from a particular form of the multivariate Laplace distribution; and,
    \item a validation that that the general form of the conjectured general solution to such multivariate asset allocation problems is not unreasonable, even though some of the specific details were incorrect.
\end{enumerate}
It is the author's understanding that neither \eqnref{laplaceholding} nor \eqnref{mvlexact2} are well known in the quantitative finance community and their sincere belief that they should be. This work illustrates that a failure of repeated mean-variance optimization, a strategy that is exactly correct when Normally distributed returns are considered, leads to overallocation to risk when asset returns in the real world are drawn from a generally more leptokurtotic distribution.
%
%
\bibliographystyle{plain}
\bibliography{citation}
\end{document}